\documentclass[10 pt,final,journal,letterpaper,oneside,doublecolumn]{IEEEtran}
\usepackage{multicol}  
\usepackage{cite} 
\usepackage{multirow}
\usepackage[pdftex]{graphicx}
\usepackage[lined,boxruled]{algorithm2e}
\usepackage{algorithmic}
\usepackage[small]{caption}
\usepackage{float}
\usepackage[utf8]{inputenc}
\usepackage{amsmath, amsthm, amssymb}

\usepackage[english]{babel}
\usepackage{amsthm}
\usepackage{xspace}
\usepackage{fancyhdr}
\usepackage{amssymb,amsmath}
\usepackage{subfig}
\usepackage{textcomp} 
\usepackage{epstopdf}
\usepackage{bbding}
\usepackage{color}
\usepackage{pifont}

\begin{document}
\title{\LARGE Opportunities for Physical Layer Security in UAV Communication Enhanced with Intelligent Reflective Surfaces}

\author{Wali Ullah Khan, \textit{Member, IEEE,} Eva Lagunas, \textit{Senior Member, IEEE,} Zain Ali, Muhammad Awais Javed, \textit{Senior Member, IEEE,} Manzoor Ahmed, Symeon Chatzinotas, \textit{Senior Member, IEEE,} Bj\"orn Ottersten, \textit{Fellow, IEEE,} and Petar Popovski, \textit{Fellow, IEEE}\thanks{This work was supported by by Luxembourg National Research Fund (FNR) under the CORE projects RISOTTI, grant C20/IS/14773976 and 5G-Sky, grant C19/IS/13713801.

Wali Ullah Khan, Eva Lagunas, Symeon Chatzinotas, and Bj\"orn Ottersten are with the Interdisciplinary Center for Security, Reliability and Trust (SnT), University of Luxembourg, 1855 Luxembourg City, Luxembourg (e-mails: \{waliullah.khan, eva.lagunas, symeon.chatzinotas, bjorn.ottersten\}@uni.lu).

Zain Ali is with Department of Electrical and Computer Engineering, University of California, Santa Cruz, USA (email: zainalihanan1@gmail.com).

Muhammad Awais Javed is with the Department of Electrical and
Computer Engineering, COMSATS University Islamabad, Islamabad 45550,
Pakistan (email: awais.javed@comsats.edu.pk)

Manzoor Ahmed is with the College of Computer Science and Technology, Qingdao University, Qingdao 266071, China (email: manzoor.achakzai@gmail.com).

Petar Popovski is with the Department of Electronic Systems, Aalborg University, Denmark (email: petarp@es.aau.dk).

\noindent{\color{red}This paper is accepted for publication in IEEE Wireless Communications Magazine.}

}}%

\markboth{Accepted for Publication in IEEE Wireless Communications Magazine}%
{Shell \MakeLowercase{\textit{et al.}}: Bare Demo of IEEEtran.cls for IEEE Journals} 

\maketitle

\begin{abstract}
%Unmanned Aerial Vehicles (UAVs) are important component of future wireless connected networks that can extend coverage in critical missions and support high data rate communication. It also provide high mobility, on demand low cast deployment and inherent line-of-sight communication.
Unmanned Aerial Vehicles (UAVs) are an important
component of next-generation wireless networks that can assist
in high data rate communications and provide enhanced coverage.Their high mobility and aerial nature offer deployment
flexibility and low-cost infrastructure support to existing cellular
networks and provide many applications that rely on mobile
wireless communications. However, security is a major challenge
in UAV communications, and Physical Layer Security (PLS) is
an important technique to improve the reliability and security
of data shared with the assistance of UAVs. Recently, Intelligent
Reflective Surfaces (IRS) have emerged as a novel technology
to extend and/or enhance wireless coverage by re-configuring
the propagation environment of communications. This paper
provides an overview of how the IRS can improve the PLS of UAV
networks. We discuss different use cases of PLS for IRS-enhanced
UAV communications and briefly review the recent advances in
this area. Then based on the recent advances, we also present
a case study that utilizes alternate optimization to maximize the
secrecy capacity for IRS-enhanced UAV scenario in the presence
of multiple Eves. Finally, we highlight several open
issues and research challenges to realize PLS in IRS-enhanced
UAV communications.
\end{abstract}


\IEEEpeerreviewmaketitle
%%%%%%%%%%%%%%%%
\begin{figure*}[!t]
\centering
\includegraphics [width=0.8\textwidth]{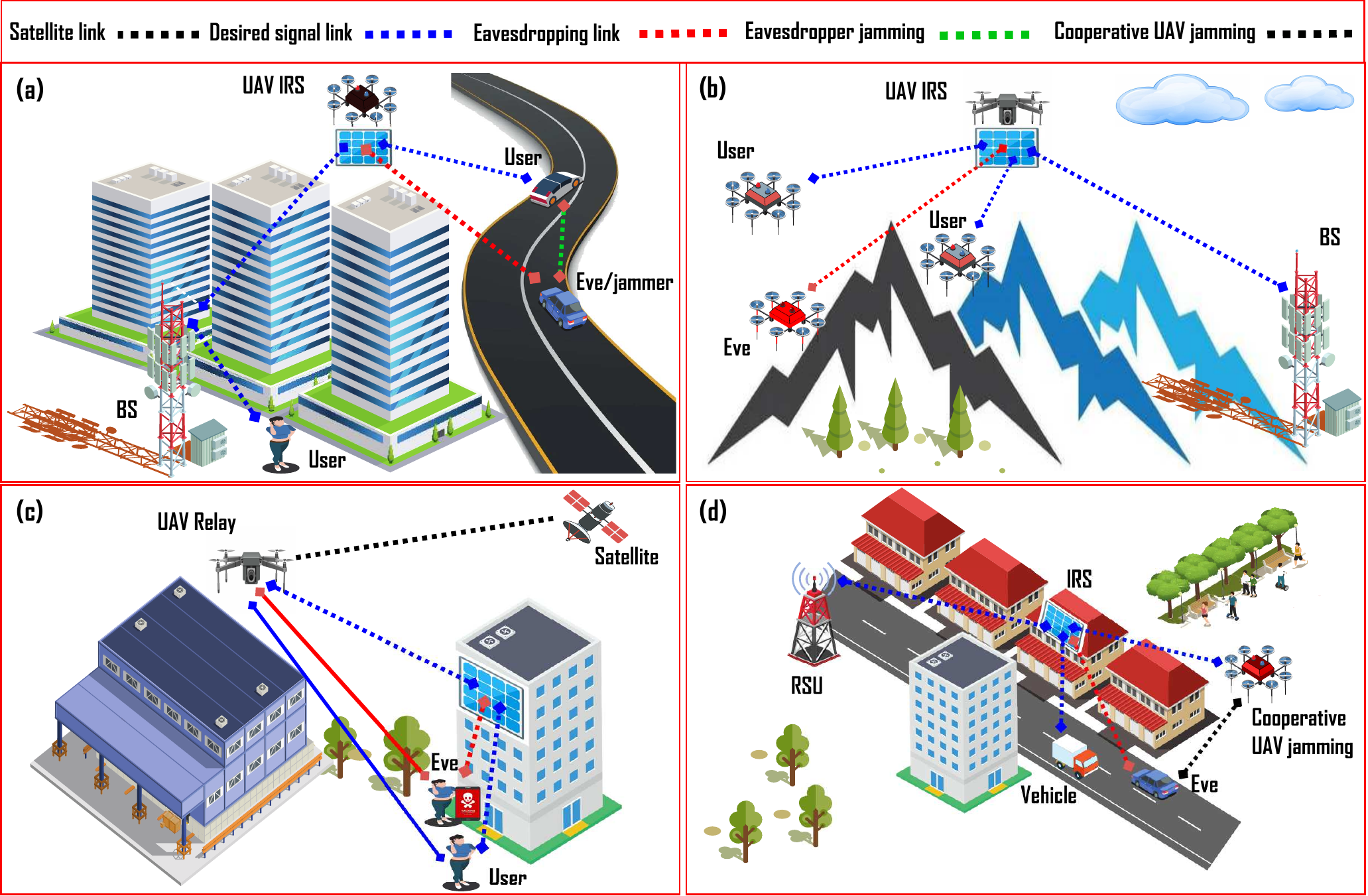}
\caption{Different cases of IRS-enhanced UAV Communication.}
\label{Fig3}
\end{figure*}
%%%%%%%%%%%%% 
\section{Introduction}
The great demand for high data rates, massive connectivity and protection from impending security attacks challenge next-generation wireless communication systems. In this regard, Unmanned Aerial Vehicles (UAVs) can play a vital role in supporting reliable and secure communications without infrastructure coverage. UAVs provide several benefits in terms of cost-friendly rapid infrastructure deployment in low signal coverage zones, mobile relay nodes to enhance coverage range, communication access points in emergency areas, and enabling Physical Layer Security (PLS) \cite{fotouhi2019survey}. Compared to traditional terrestrial communications, UAVs provide strong channel conditions due to line of site transmission links. Therefore, there is high demand for deploying large number of UAVs to support next-generation wireless networks \cite{uav_1}. UAVs can be used for different applications in transportation systems, cellular communications, agriculture, and emergency management. In addition, the use of UAVs will be an integral part of next-generation wireless networks to support ultra-low latency, and extreme reliability applications \cite{uav_2}. 

Despite the promising features, several challenges exist in deploying UAVs for next-generation wireless networks. Among different challenges, energy efficiency, security, and reliability are the major challenges in developing efficient UAV communications. Due to limited energy reserves on board, intelligent energy usage and replenishment mechanisms are required for energy-aware UAVs deployments and operations. Besides, with the increase in malicious attacks on static and mobile networks, it is critical to use robust Cryptographic algorithms to mitigate them. However, more complex Cryptographic algorithms entail significant overhead, thus increasing the packet size. It also significantly increases the required bandwidth and puts more transmission burden on the available spectrum. As a result, transmission reliability is compromised, and network capacity is reduced.   

Cryptography-based information confidentiality seems inapplicable for next-generation technologies and networks for the following reasons: Cryptography algorithms are based on computationally hard problems, and today's adversaries have unlimited computing resources support and can be disastrous for crypto-systems \cite{8332096}. In addition, conventional centralized key sharing and management processes for distributed networks like IoT, UAVs, and vehicular networks are challenging. Moreover,  significant communication overhead is caused by complex upper-layer operations and can increase the cost and complexity of user equipments/devices. To overcome the challenges faced by traditional cryptographic security techniques, Physical Layer Security (PLS) is a valuable technique that can work in conjunction with cryptographic techniques to secure UAV communications while maintaining the reliability of transmissions \cite{sun2019physical}. PLS uses the randomness of the wireless medium to prevent transmissions from eavesdropping. However, as PLS-based UAV communications rely on the strength of signal transmissions between users and UAVs, their performance is degraded in low signal coverage areas and non-line of sight scenarios \cite{irs_1}.

Intelligence Reflective Surface (IRS) has recently emerged as a emerging technology due to its high energy efficiency. IRS is very effective for secure coverage extension in non-line-of-sight communication scenarios. It can efficiently assist the signal delivery if the transmitter and receiver do not have a direct link. IRS consists of a large number of passive reflecting elements which can intelligently reconfigure the signal direction toward the receiver. According to the IRS principle, the phase of the incident signal from the transmitter can be smartly shifted towards the receiver without consuming any energy. IRS has recently been integrated into UAV networks for energy-efficient and secure communications. IRS can improve the PLS of UAV communications by constructively adding the signal to the user and destructively to the Eve \cite{wang2020intelligent,irs_1, irs_2}. As a result, the secrecy performance of PLS in many UAV-to-UAV and UAV-to-user communication scenarios can be improved using IRS. However, a major challenge that hinders widespread adoption of PLS is the difficulty of accurately detecting Eves and their locations. As UAVs can be equipped with cameras and sensors in such a way that they can map the environment and detect potential Eve.

This paper provides an overview of IRS-enhanced PLS for UAV communications. We present four major use cases of IRS-enhanced PLS in UAV communications related to improving secrecy rate in non-LOS scenarios, satellite communications, mobile IRS-enhanced UAVs, and cooperative jamming. We also discuss the recent work in the literature related to IRS-enhanced PLS for UAV communications. Moreover, we present a UAV communications-based case study highlighting the significance of IRS for  PLS. Further, we propose an alternate optimization-based algorithm to maximize the secrecy rate of UAV communications in the presence of IRS. Finally, we highlight several open research challenges to realize IRS and PLS-based UAV communications.

% PLS use cases in the context of I-UAVS and present significant challenges. We also present a case study simulating the PLS-assisted I-UAVs scenario and show improvement offered by this synergy in terms of sum rate. Finally, we present several critical challenges to improve PLS-assisted I-UAVs, including PLS performance against malicious attacks, channel modeling of I-UAVs, machine learning-based techniques, and RIS for PLS-assisted I-UAVs.

\section{IRS-Enhanced PLS for UAV Communication: Overview, Use Cases, and Recent Advances}

This section provides an overview of IRS-enhanced PLS for UAV communications and highlights several use cases and recent work in this area. 

\subsection{Overview of IRS-Enhanced PLS for UAV Communications}

PLS leverages wireless channel characteristics to improve the information received at the legitimate destination compared to Eve. There are different techniques to achieve efficient PLS, including coding, signal processing, transmission power control, and jamming techniques \cite{sun2019physical, bjorn_pls}. PLS can provide efficient and secure key generation, authentication, and defense against Eves. A significant advantage of using PLS for security is its low cost of computation and packet sizes. PLS is thus particularly useful for energy-constrained devices such as UAVs and IoT nodes. 

IRS is composed of meta-surfaces that can reflect an incoming signal towards the desired destination. A key feature of IRS is its reconfigurability which allows controlling the phase shifts of its elements so that signal reception at the destination can be maximized. IRS can be placed as stand-alone surfaces or installed at buildings and placed in areas where signal coverage is low. IRS can be beneficial in non-line-of-sight communication scenarios, improving the reliability of communications \cite{irs_1, irs_2}. 

IRS can play a vital role in improving information confidentiality of UAV communications by providing an enhanced secrecy rate. In fact, UAVs can also be furnished with IRS and assist in secure communication in low coverage areas on the fly. With the help of reconfigurable reflecting elements, IRS can significantly improve the PLS of UAV communications. Specifically, IRS adds the incident wireless signal constructively to the user receiver but destructively to the Eve. Moreover, it can also act as a green jammer to attack Eve by producing jamming signals without consuming external energy. Besides security problems, IRS-assisted UAVs can provide reliable and energy-efficient communications, thus improving the battery life of UAVs.

\begin{table*}[!htp]
\renewcommand{\arraystretch}{1.5}
 \caption{Recent advances in IRS-enhanced PLS for UAV communications}
 \label{tab:t1}
  %\resizebox{\textwidth}{!}{%
\begin{tabular}{llllll}
\hline \hline
\textbf{PLS aspect}   & \textbf{IRS aspect}                                                            & \textbf{UAV aspect}                                                            & \textbf{Scenario}                                                                            & \textbf{Technique}                                                                                                 & \textbf{Results}      \\ \hline
Maximize secrecy rate \cite{lit_1} & Phase shift control                                                            & \begin{tabular}[c]{@{}l@{}}Trajectory control\\ Power control\end{tabular}     & \begin{tabular}[c]{@{}l@{}}UAV to single receiver\\ Single Eve\end{tabular}         & Alternate optimization                                                                                             & Improved secrecy rate \\ \hline
Maximize secrecy rate \cite{lit_2} & \begin{tabular}[c]{@{}l@{}}Beamforming design\\ Position design\end{tabular}   & \begin{tabular}[c]{@{}l@{}}Beamforming design\\ Position design\end{tabular}   & \begin{tabular}[c]{@{}l@{}}UAV BS to receiver\\ Single Eve\end{tabular}             & \begin{tabular}[c]{@{}l@{}}Ideal beamforming model\\ Alternate optimization\\ Semidefinite relaxation\end{tabular} & Improved secrecy rate \\ \hline
Maximize secrecy rate \cite{lit_3} & Phase shift control                                                            & \begin{tabular}[c]{@{}l@{}}Power control\\ Position design\end{tabular}        & \begin{tabular}[c]{@{}l@{}}UAV to ground user\\ Single Eve\end{tabular}             & \begin{tabular}[c]{@{}l@{}}Fractional programming\\ Alternate optimization\end{tabular}                            & Improved secrecy rate \\ \hline
Maximize secure EE \cite{lit_4}    & \begin{tabular}[c]{@{}l@{}}Phase shift control\\ User association\end{tabular} & \begin{tabular}[c]{@{}l@{}}Power control\\ Trajectory design\end{tabular}      & \begin{tabular}[c]{@{}l@{}}BS to users\\ IRS-enhanced UAV\\ Single Eve\end{tabular} & \begin{tabular}[c]{@{}l@{}}Linear programming\\ SCA\\ Alternate optimization\end{tabular}                          & Improved secure EE    \\ \hline
Maximize secrecy rate \cite{lit_5} & Beamforming design                                                             & Trajectory design                                                              & \begin{tabular}[c]{@{}l@{}}UAV to ground user\\ Single Eve\end{tabular}             & \begin{tabular}[c]{@{}l@{}}SCA\\ S-procedure\\ Semidefinite relaxation\\ Alternate optimization\end{tabular}       & Improved secrecy rate \\ \hline

Maximize secrecy rate \cite{lit_6} & Phase shift control                                                               & Position design & \begin{tabular}[c]{@{}l@{}}BS to users\\ Single Eve\end{tabular}          & Iterative algorithm                                                                 & Improved secrecy rate \\ \hline
                                                         
\end{tabular}
%}
\end{table*}

\subsection{Use Cases of IRS-Enhanced PLS for UAV Communications}
IRS can provide multiple benefits to improve PLS working in UAV networks. We discuss a few of these use cases in this subsection as highlighted in Fig.~\ref{Fig3}.

\subsubsection{Mobile IRS-Enhanced UAVs}
The mobility of a UAV can be used in collaboration with the intelligent signal reflection feature of IRS to enhance ground-to-ground and air-to-air communications, as presented in Fig.~\ref{Fig3} (a). IRS can be placed on a UAV to improve the communications and security of vehicular networks (for vehicle-to-vehicle communication). The BS can identify the areas on the road where the secrecy rate is low based on the channel quality values received from different vehicles and Eves. The mobile IRS can be directed to areas where PLS is compromised and acts as a relay to improve the desired signal and the secrecy rate.

\subsubsection{Improved Secrecy Rate in Non-LOS Scenarios}
UAV communication faces challenges of difficult terrain and signal blockages. As shown in Fig.~\ref{Fig3} (b), two UAVs connected in a mountainous area can face signal disconnections, making PLS less effective. An eavesdropper UAV can take advantage of this scenario as the secrecy rate will be reduced. Similarly, in air-to-ground communication between a UAV and a base station, such line of sight blockages can reduce the security levels. IRS can be very effective in the above scenarios as it can improve the line of sight communications in case of signal blockages. IRS can be installed at strategic locations such as buildings, or a mobile UAV IRS can be used. IRS will improve the rate between the two UAVs or between the UAV and BS, thus improving the secrecy rate.        

\subsubsection{Satellite Communications}
Satellite communications use IRS as a relay for transmitting signals to the ground base station as described in Fig.~\ref{Fig3} (c). Generally, the downlink communication from UAV to BS suffers attenuation and fading. Therefore, IRS can be installed between the path of UAV and BS, i.e., on buildings, thus facilitating better communication of BS with the satellite. Similarly, UAVs equipped with IRS can be configured such that satellite signals can reach the desired destination with high reliability.

\subsubsection{Cooperative Jamming}
Eavesdropper jamming signals are a significant threat to PLS in UAVs. A sample scenario is shown in Fig.~\ref{Fig3} (d), where an Eve sends jamming signals to the user. By using IRS, the signal-to-noise ratio of the desired signal can be further strengthened. In addition, intelligent beamforming at IRS can decrease the signal strength of the actual message received by the Eve. Moreover, cooperative jamming in which BS generates an artificial noise and IRS directs it to the Eve can mitigate the impact of jamming attacks by the Eve.

\subsection{Recent Advances in IRS-Enhanced PLS for UAV Communication}

Little work has been reported related to IRS-enhanced PLS for UAV communications. The work in \cite{lit_1} proposes physical security and an IRS-enhanced UAV framework. The scenario considers UAV transmission to the desired destination node in the presence of an Eve that can intercept the message. The work aims to jointly optimize the UAV's transmit power and trajectory and control IRS's phase shift to maximize the secrecy rate using an alternate convex approximation algorithm. The algorithm selects an initial transmit power and UAV trajectory values to obtain an optimal solution and calculates the IRS phase shift value. Then, the secrecy rate for the above parameters is also evaluated. The transmit power and trajectory are evaluated alternately till the algorithm converges and provides the maximum secrecy rate. Results show that the secrecy rate of the IRS-enhanced UAV network is maximized.
In \cite{lit_2}, the authors consider a mmWave network in which a single UAV BS transmits the message to the desired destination node in the presence of a single Eve. The paper aims to design the position and beamforming of both UAV BS and IRS to maximize the secrecy rate. The paper assumes an ideal beamforming model in which the signal is received only by the desired receiver, and the Eve cannot intercept the message. The proposed algorithm is split into two phases. In the first phase, the UAV and the IRS positions are designed to maximize the secrecy rate. An alternate optimization technique is used in the second phase to optimize the UAV BS beamforming vector and IRS passive beamforming. Finally, a semidefinite relaxation technique is used to reduce the complexity of the optimization. As a result, the proposed protocol achieves a higher secrecy rate than the other techniques in the literature.   

The work in \cite{lit_3} proposes an algorithm to maximize the secrecy rate for UAV to ground user transmissions in the presence of a single Eve. The optimized parameters to achieve the above goal include the UAV's transmit power,  location, and the IRS's phase shift. At first, the transmit power is selected for a fixed UAV location and IRS phase shift. Then, the UAV location problem is solved using the convex algorithm's difference, and the IRS phase shift is optimized using fractional programming. Finally, an alternate optimization is applied to find the optimal values of three parameters. The secrecy rate of the proposed technique has been shown to outperform other baseline algorithms. 

In \cite{lit_4}, the authors aim to maximize the network's secure Energy Efficiency (EE) consisting of a BS,  an IRS-equipped UAV, multiple users, and a single Eve. EE is the value of the minimum secrecy rate per total power used by the network. The user association problem is relaxed using continuous variables and solved via linear programming. Further, the user association problem's integer output is obtained using the rounding technique. The power control is optimized using the Successive Convex Approximation (SCA) method. The trajectory design and phase shift control problems are solved using alternate optimization. Simulation results verify the improvement in the secure EE of the network. 

The work in \cite{lit_5} considers a UAV and ground user communication scenario assisted by an IRS in the presence of a single Eve. A time division multiple access technique is used for uplink and downlink communication. In addition, the work considers an imperfect channel state information. The proposed algorithm jointly optimizes the UAV trajectory, beamforming of IRS, and users' transmit power. To solve these problems, three techniques, namely SCA, S-procedure, and semidefinite relaxation, are used along with alternate optimization. Performance evaluation of the proposed algorithm highlights the significance of IRS in improving PLS and secrecy rate.  

In \cite{lit_6}, the authors propose an iterative algorithm to maximize the secrecy rate for BS to user communications in the presence of a single Eve. The proposed algorithm controls the phase shift of IRS and the UAV's position to improve the transmissions' secrecy rate. The role of the UAV is to act as a passive relay and facilitate the physical layer security.

Most of the works in the literature consider a single Eve for PLS and IRS-enhanced UAV communications. Moreover, they also assume that the location/position of the user and Eve is static. In the next section, we present a more practical case study highlighting the advantages of using IRS when multiple Eves are considered in the vehicular network, i.e., mobile users and Eves. 

%In \cite{lit_6}, authors proposed an algorithm to maximize the secrecy rate between users and UAVs in the presence of multiple Eves. A mmWave communication scenario is considered, and imperfect channel state information is assumed. A double Deep Deterministic Policy Gradient (DDPG) based Deep Reinforcement Learning (DRL) method is used to reduce the complexity of the solution for the considered problem. The UAV active beamforming and IRS passive beamforming are solved using the first DDPG, whereas the UAV trajectory is optimized using the second DDPG method. Results show that the proposed DRL method maximizes the secrecy rate compared to the other learning techniques.    

%%%%%%%%%%%%%%%%
\begin{figure}[!t]
\centering
\includegraphics [width=0.45\textwidth]{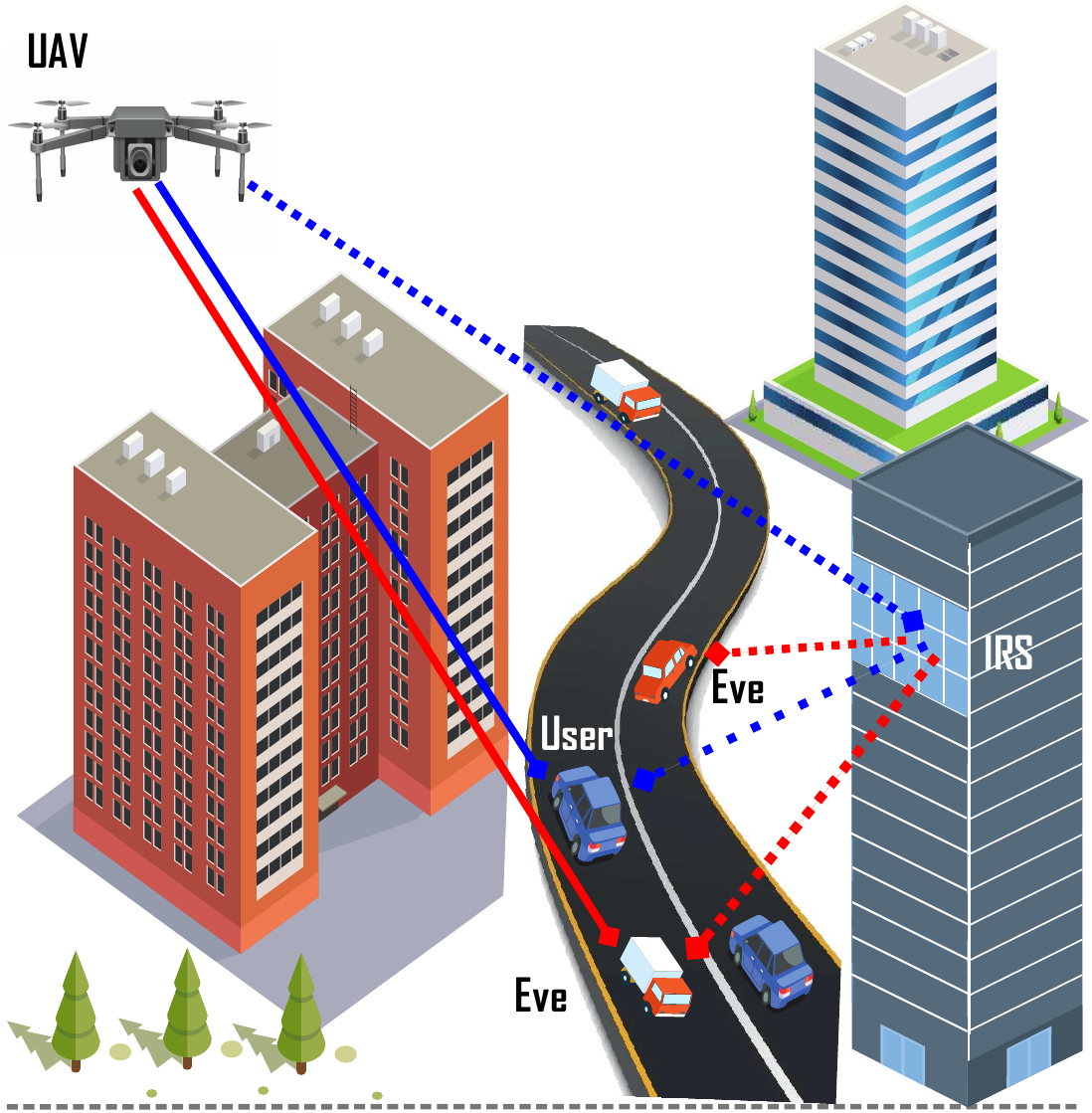}
\caption{System model}
\label{Fig3}
\end{figure}
%%%%%%%%%%%%% 

\section{Optimizing PLS of IRS-Enhanced UAV Communications: A Case of Multiple Eavesdroppers}
As shown in Fig. \ref{Fig3}, we consider a communication system where a UAV communicates with a legitimate vehicle on the road in the presence of $K$ non-colluding Eve vehicles. The UAV and vehicles are equipped with a single antenna scenario. To improve the channel capacity, the considered system model also consists of an IRS with $M$ passive elements mounted on the top of the building. Thus, the vehicle receives a signal from a UAV through a direct link and an IRS-assisted link. Further, it is assumed that the transmitted signal is received by both: the legitimate vehicle and the Eve vehicles through direct and indirect links. We aim to maximize the system's secrecy capacity subject to the constraints of UAV battery capacity and the phase designing matrix of IRS passive component, where the secrecy capacity is defined as the difference in the rates of the legitimate vehicle and the Eve vehicle with maximum signal to interference plus noise ratio.

We formulate the secrecy capacity maximization problem subject to the UAV battery capacity and the IRS passive component. The considered problem is non-convex, where obtaining the global optimal solution is challenging. However, alternate optimization techniques can be employed to solve such problems efficiently. Alternate optimization provides an efficient way of decoupling the problem into the optimization variables. Then, the problem is solved for one variable at a time, while alternating between the different variables multiple times. This provides an effective method to find the efficient joint solutions for all the variables while considering a single variable at a time. Therefore, an alternate optimization-based gradient descent optimizer is adopted to find the suboptimal yet efficient solution to the formulated secrecy capacity maximization problem.

For the results, we adopt a Monte Carlo simulation where 10,000 independent experiments were conducted, and the figures show the average performance of the scheme. Further, we consider independently and identically distributed Rayleigh fading channels drawn from complex Normal distribution with unity variance and 0 mean. Unless stated otherwise the simulation parameters are set as follow. The transmit power of UAV is 3 Watts, the number of IRS passive elements is 10, the number of non-colluding Evs vehicles is 8, path-loss exponent is 3, height of UAV is 80 meters, and noise variance is 0.01. Moreover, we define a passive reflection matrix of IRS as ${\Theta}=\text{diag}\{\varphi_{1}e^{j\theta_{1}}, \varphi_{2}e^{j\theta_{2}}, \dots, \varphi_{M}e^{j\theta_{M}}\}$, where $j=\sqrt{-1}$, $\varphi_{m}\in[0,1]$ denotes the amplitude and $\theta_{m}\in[0,2\pi]$ is the phase shift of element $m$. Finally, we compared two scenarios: IRS-enhanced UAV communication and conventional UAV communication without IRS. This comparison aims to highlight the advantage of using the IRS to enhance the system's secrecy capacity. 

To ensure positive secrecy capacity at all times, many works in the literature considered that the channel's gain from the transmitter to Eve is always less than the channel gain from the transmitter to the user. Although this assumption is reasonable in many cases, it is not always possible to have a better channel to the user than the channels between the transmitter and Eve. Therefore, we consider a more practical scenario where the channels are independent in the true sense, and Eve can have a better channel than the user. 
%%%%%%%%%%%%%%%%
\begin{figure}[!t]
\centering
\includegraphics [width=0.48\textwidth]{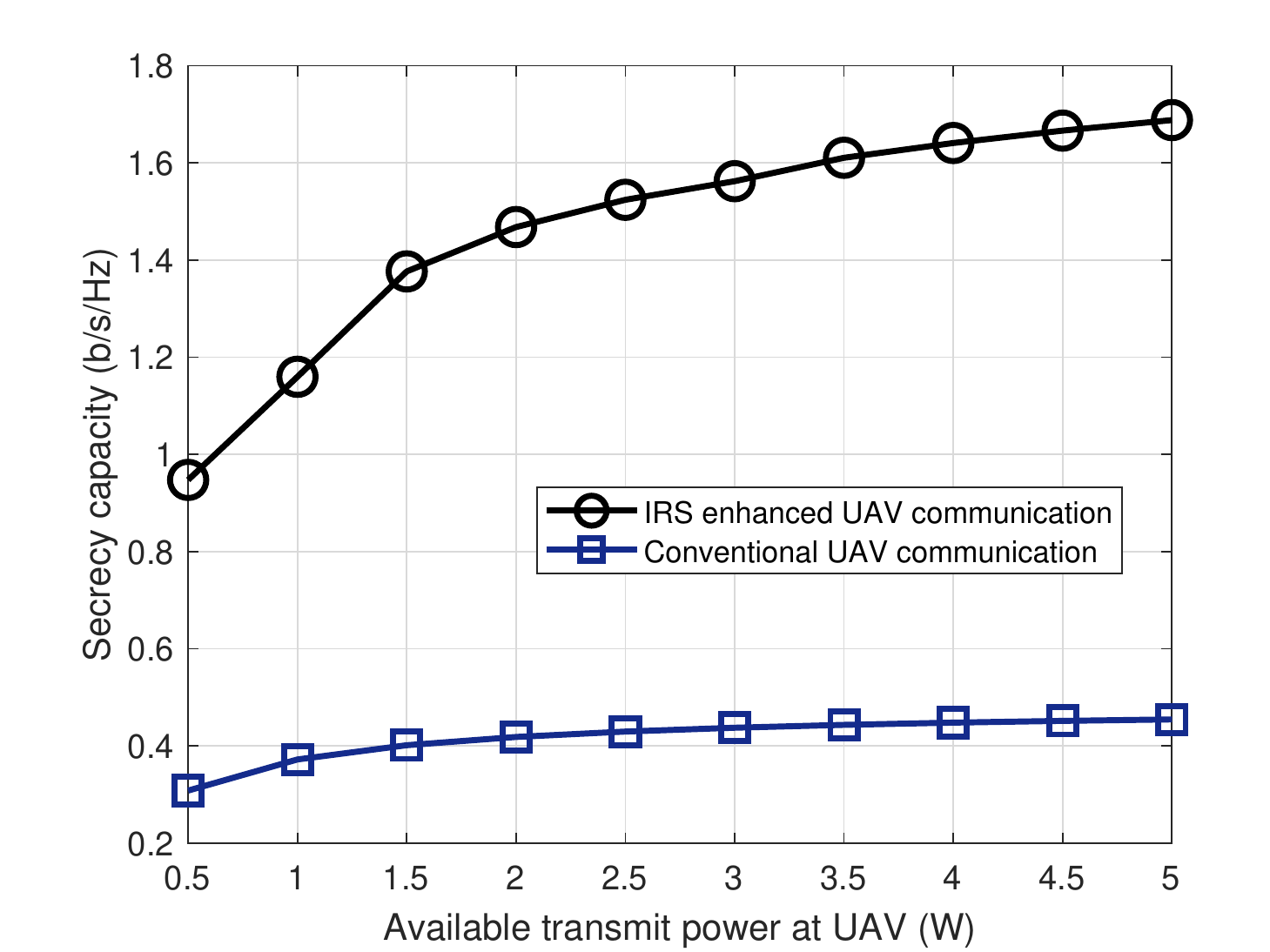}
\caption{The figure shows the impact of increasing available transmission power at the UAV on the secrecy capacity of the system. This figure shows the advantage of using IRS in communication systems clear.}
\label{Figr1}
\end{figure}
%%%%%%%%%%%%% 

\begin{figure}[!t]
\centering
\includegraphics [width=0.48\textwidth]{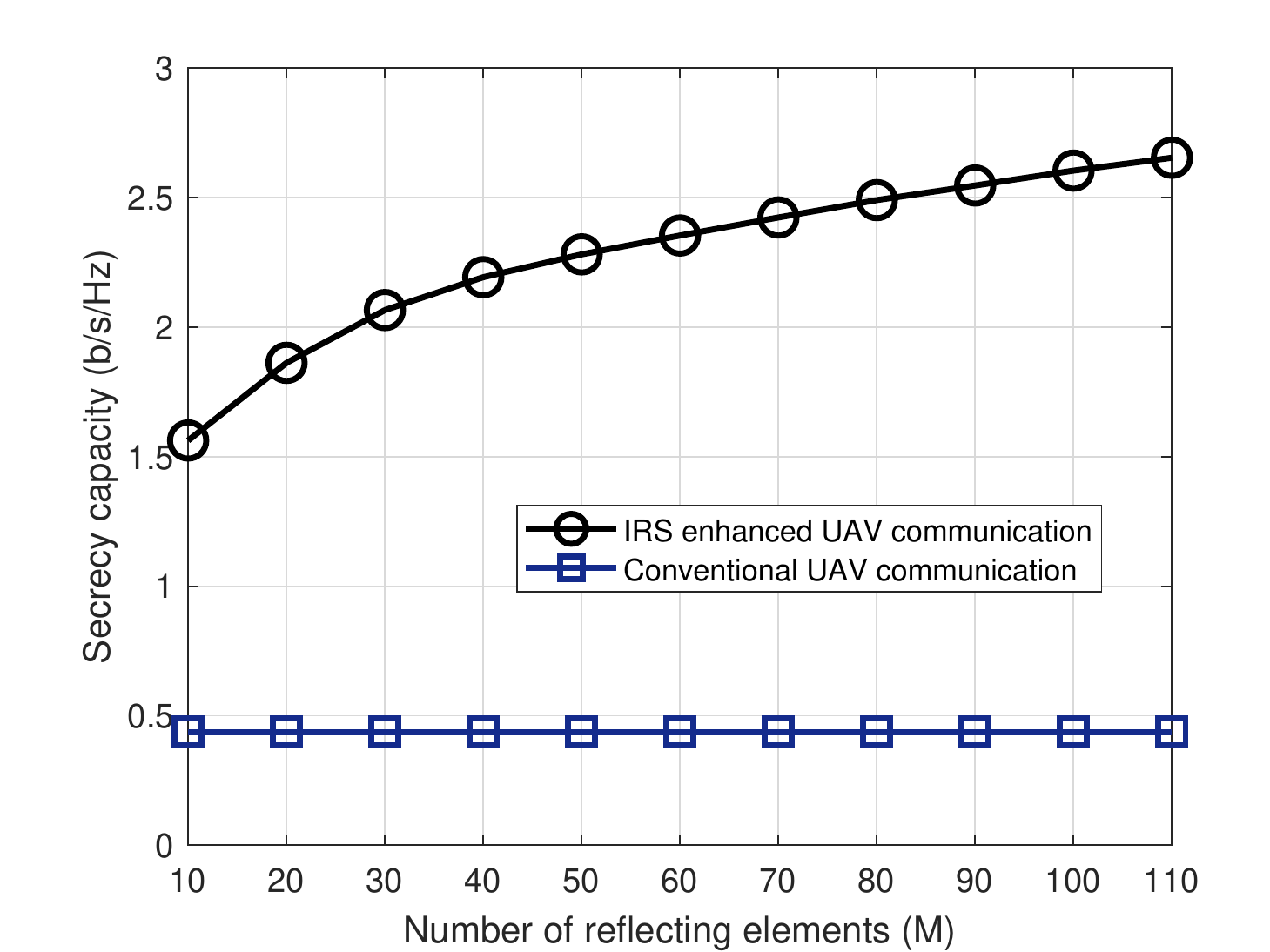}
\caption{This figure shows that increasing the number of reflecting elements of the IRS improves the secrecy capacity of the system.}
\label{Figr2}
\end{figure}

In the system without IRS, it was observed that if the channel conditions at the Eve vehicle are better than the optimization results in a 0 W transmit power. However, as in this case, it is impossible to have a greater rate value at the legitimate vehicle than the Eve vehicle. Hence, to avoid a negative value of secrecy capacity, the system decides not to transmit, resulting in a 0 secrecy capacity. However, when the system is equipped with an IRS, even if the channel from the UAV to Eve vehicles is better than the channel between UAV and legitimate vehicle, the IRS elements can be adjusted to achieve positive secrecy capacity. Thus, a significant gain in the secrecy capacity is obtained with IRS, as shown in Fig. \ref{Figr1} and Fig. \ref{Figr2}.

For the results in Fig. \ref{Figr1}, we considered ten reflecting elements at the IRS and three Eve vehicles. Figure \ref{Figr1} shows that when the UAV's available power increases, the system's secrecy capacity also increases. However, the increase in secrecy is logarithmic as at the lower values of available power, a more significant gain in secrecy is observed when power is increased. Further, with the increase in power, the gap in the secrecy provided by IRS and non-IRS systems also increases. Similarly, for the results in Fig. \ref{Figr2}, we considered the available power to be 3 W, and the number of Eve vehicles is also 3. The figure shows that increasing the number of reflecting elements in the IRS improves the secrecy performance of the system, and the gap in the performance of IRS and non-IRS systems also increases. However, just as in the previous case, when the number of reflecting elements increases, the gain in secrecy is logarithmic.
%%%%%%%%%%%%%%%%
%%%%%%%%%%%%% 

\section{Open Issues and Future Research Directions}
IRS-enhanced UAV Communication is an emerging area
of research, and dealing with the security challenges that
come with it is key to achieving its full potential. We list several open issues
and challenges for future development and work.
\subsection{PLS against Malicious IRS-Enhanced UAV Attacks}
High mobility and flexibility are essential in the
case of IRS-enhanced UAV communication to improve the PLS, intercept sensitive information, and even jam the legitimate links of UAVs to decrease their
quality. In particular, when a malicious UAV is equipped with IRS and can intercept the information of other UAVs. In that case, safeguarding the communication of UAV systems could provide
more challenges than dealing with traditional terrestrial
Eve. However, no research works
have been performed on this crucial topic from the perspective
of communication theory. Therefore, it is preferred to
explore advanced methods in terms of PLS to protect against
malicious IRS-enhanced UAV attacks. 
\subsection{PLS against Pilot Contamination Attacks}
Effective beamforming requires accurate CSI for secure
IRS-enhanced UAV communication could be obtained by utilizing the pilot signals. However, in some cases, intentional deterministic pilot samples are sent by active Eve, similar to those transmitted by legitimate transmitters to deceive the UAV and facilitate eavesdropping.
As a result, the UAV formulates an ineffective transmission
technique that can benefit the signal reception of malicious Eves. For example, although information
leakage is possible for confidential data delivery, the UAV may misunderstand the network environment and fly too close to the Eve.
Thus, finding some efficient guidelines to minimize the impacts of pilot contamination attacks is complex but
essential for IRS-enhanced UAV safety.
\subsection{Cooperative Jamming for IRS-Enhanced UAV Communications}
The IRS-enhanced UAV can act as a friendly jammer to protect the user data by sending artificial noise toward the malicious attacker. In this way, the PLS of a system can improve; however, it requires additional energy consumption of energy constraint UAVs. It is important to note that energy efficiency is crucial for UAVs because of their limited onboard energy reservoirs, which becomes a bottleneck and affects the performance of UAVs. Thus, energy efficiency solutions that reduce the total energy consumption without affecting the system performance need to be further investigated. Moreover, IRS mounted over the UAVs can be used to tune the phase shift of the signals smartly. As a result, the original and reflected signal adds constructively to the legitimate user to enhance signal-to-noise ratios.
Further, IRS can also be used as a friendly jammer to reduce the effect of Eves. For instance, IRS can use different phase shifts to produce destructive signals to reduce signal strength in any specific direction and decrease eavesdropping chances. In this regard, more research on secure communication techniques is needed when combining IRS and UAVs.
\subsection{Artificial Intelligence/Machine Learning Approaches for IRS-Enhanced UAV Communication}
Optimizing large-scale IRS-enhanced UAV communications is challenging, especially when UAVs are deployed in a partially unknown environment. In particular, optimizing UAV trajectory, IRS reflecting elements, and resource optimization of the entire network is challenging to design due to non-linear models. Thus, designing approaches with low complexity and efficient system performance is challenging. Artificial intelligence/learning approaches are powerful tools for designing and optimizing such networks. These approaches are rapidly developing and provide promising and robust tools for designing and optimizing challenging scenarios.
Moreover, the hybrid models, data-driven methods, and hybrid offline and online methods to improve secrecy performance can be used to analyze the complex system. However, several challenges still need to be investigated—for example, large computational processing power, high energy consumption, and latency.

\subsection{Key Size Reduction of Cryptographic Algorithms}

PLS can be useful to reduce the size of transmitted data by reducing the reliance on complex Cryptographic algorithms. Furthermore, PLS can work in conjunction with Cryptographic algorithms to improve the level of security at a low transmission cost. For example, the Elliptic Curve Digital Signature Algorithm (ECDSA) with a reduced key size may be used for UAV communications in case PLS also provides defence against Eves. The advantage of using low-key size algorithms is twofold. One is the reduced message size and improved use of the available spectrum. The other advantage is reduced latency as transmission delay and signature verification delay are reduced. In the future, there is a need to investigate the optimal key sizes which may be sufficient for different UAV application scenarios.

\subsection{Resource Optimization for IRS-Enhanced UAV Communications}
Optimal resource allocation is essential in wireless communications to enhance performance. Compared to conventional UAV communication, IRS-enhanced UAV communication adds a new dimension to improve PLS by optimizing IRS phase shift of passive elements and beamforming. Since IRS and resource allocation (i.e., trajectory design, sub-carrier, power allocation, etc.) are frequently connected, the design optimization issue becomes intractable, and the existing designs are sub-optimal. However, the performance variances between optimal and the current sub-optimal solutions are not apparent. Therefore, optimal techniques are needed to enhance the PLS of IRS-enhanced UAV communication in different applications while balancing computational complexity and system performance. 

\subsection{Availability of CSI in IRS-Enhanced UAV Communications}
The recent studies on PLS of IRS-enhanced UAV communication assume the availability of perfect CSI at the transmitter and/or the IRS. However, estimating the channel for the IRS-enhanced UAV system is challenging because of the large number of passive elements. More specifically, these elements are passive in nature without signal processing capabilities. Thus, they do not have active transmitting and receiving abilities. Based on the above observation, the transmitter can achieve imperfect CSI in practice. Another critical point is to note that the CSI of the legitimate user is only available when it is active or registered with the network. However, the CSI of the passive Eve is not available. 

\section{Conclusion}
This paper provides an introduction to IRS-enhanced PLS for UAV communications. The work discusses the major use cases of IRS to improve the PLS working in UAV communications. We discuss recent works in this area, mainly related to UAV and ground user communications in the presence of a single Eve. We also present a case study that maximizes the secrecy rate of UAV communications in the presence of multiple Eves. The proposed work uses alternate optimization to control the phase shift of IRS and the UAVs' trajectory to enhance UAV communications. Simulation results show that IRS significantly improves the secrecy rate of UAV communications. We also discuss future research challenges related to IRS and PLS in UAV communication scenarios.

\ifCLASSOPTIONcaptionsoff
  \newpage
\fi

%\bibliographystyle{IEEEtran}
\bibliographystyle{IEEEtran}% This is IEEEtran.bst file
\bibliography{Wali_EE}

\end{document}